\documentclass[10pt,onecolumn]{article}
\usepackage{epsf}
\begin{document}

\title{Final State Interactions and the Effects of Potentials 
on Particle Reactions}
\author{ 
G. Castellani, S. Reucroft, Y.N. Srivastava$^\dagger $, \\ 
J. Swain and A. Widom \\ 
{\it Physics Department, Northeastern University} \\ 
{\it Boston Massachusetts, USA}\\ 
$^\dagger $ {\it Physics Department \& INFN, University of Perugia}\\  
{\it Perugia Italy}}

\date{}
\maketitle

\begin{abstract}
In nuclear physics, it is well known that the electromagnetic
(Coulomb) interaction between final state products can drastically effect
particle reaction rates. Near thresholds, for example, nuclear alpha decay
is suppressed while nuclear beta decay is enhanced by final state Coulomb
interactions. Here we discuss high energy physics enhancement and/or
suppression of reactions wherein the potentials must include weak and strong
as well as electromagnetic interactions. Potentials due to the exchange of
gluons and the exchange of a hypothetical Higgs particle
are explicitly considered.
\end{abstract}

\section{Introduction}

The Coulomb interaction
\begin{equation}
U_{Coul}=\left(\frac{e^2}{4\pi \epsilon_0}\right)\left(\frac{Z_1Z_2}{r}\right)
=Z_1Z_2\left(\frac{\hbar c\alpha }{r}\right)
\label{intro1}
\end{equation}
between the final products of nuclear reactions can have a large effect on
particle reaction rates and cross sections. If
the final state Coulomb potential is repulsive, then the reaction is suppressed.
Such is the case for (say) nuclear alpha decay or inverse nuclear beta decay. If the
Coulomb final state interaction is attractive, then the reaction is enhanced. Such is
the case for nuclear beta decay. The effects of the final state Coulomb
potential is (i) particularly large near threshold and (ii) requires methods
far beyond standard low order perturbation theory for a proper calculation.

Although the application of final state interaction theory to problems of
nuclear physics is by now fairly routine, the theory is {\em not} yet quite
standard practice in high energy physics wherein perturbation theory perhaps
{\em too often reigns supreme}. Yet the potentials of the weak and strong
interactions, if {\em not} the gravitational potential
\begin{equation}
U_{Newton}=-G\left(\frac{M_1M_2}{r}\right),
\label{intro2}
\end{equation}
surely play a final state interaction role similar to the Coulomb interaction in
nuclear physics. In particular, we wish to discuss these final state interactions
which are derived from both weak and strong forces. There has been considerable earlier
work\cite{1,2,3,4,5,6,7,8,9,10,11,12,13} with applications to the
\begin{math} W^+W^- \end{math} and heavy flavor
\begin{math} q\bar{q} \end{math} production.

The strong force potential, presumed due to gluon exchange, has the form
\begin{equation}
U_{Glue}=\left(g^2\over 4\pi \epsilon_0 \right)
\left(\frac{{\bf T}_1\cdot {\bf T}_2}{r}\right)
=\left(\frac{\hbar c\alpha_s }{r}\right){\bf T}_1\cdot {\bf T}_2,
\label{intro3}
\end{equation}
in which the matrices \begin{math} \{{\bf T}\}  \end{math} are the color
\begin{math} SU(3) \end{math} group generators. The gluon exchange potential
Eq.(\ref{intro3}) is written down in close analogy to the photon exchange
potential Eq.(\ref{intro1}); It reads
\begin{eqnarray}
U_{\bar{q}q} &=& V_{q\bar{q}}=
-\frac{4}{3}\left(\frac{\hbar c\alpha_s }{r}\right)
\ \ \ {\rm (quark\ anti-quark)},
\nonumber \\
U_{\bar{q}\bar{q}} &=& V_{qq}=
-\frac{2}{3}\left(\frac{\hbar c\alpha_s }{r}\right)
\ \ \ {\rm (quark\ quark)}.
\label{intro3a}
\end{eqnarray}
However, Eqs.(\ref{intro3}) and (\ref{intro3a}) hold true only in the
\begin{math} r\to 0 \end{math} limit. For large \begin{math} r \end{math},
the presumed confinement (linear) portion of the potential is presently
only partially understood. The details of the full quark potentials
are summarized in \ref{qp}.

The weak Higgs exchange potential has the form
\begin{equation}
U_{Higgs}=-\left(\sqrt{2}G_F\over 4\pi \right)
\left(\frac{M_1M_2}{r}\right)e^{-(M_H c/\hbar)r }
\label{intro4}
\end{equation}
in close analogy to the graviton exchange potential Eq.(\ref{intro2}).
Here, the Fermi interaction strength \begin{math} G_{F} \end{math}
plays a role analogous to the Newtonian gravitational coupling
\begin{math} G \end{math} while the mass
\begin{math} M_H \end{math} of the Higgs particle plays the role of an
inverse screening length. That the graviton exchange potential should
bear a strong resemblance to the Higgs exchange potential (apart from
screening) is due to the fact that gravitational mass is the source and
sink of the gravitational field while inertial mass is the source and
sink of the Higgs field. The principle of equivalence between gravitational
and inertial mass dictates that the Higgs particle (if it exists) is
intimately connected with gravity.

To compute the final state interaction effects of the effective exchange
potentials which may enhance or may suppress the reaction, it is convenient
to employ the quasi-classical relativistic Hamilton-Jacobi equation. If the
potential is repulsive and the reaction suppressed, then the effect lies
mainly in the classically disallowed region (quantum tunneling). If the
potential is attractive and the reaction is enhanced, then the effect arises
due to the strong overlap of the attracted particle wave functions.
This point is illustrated in Sec.\ref{coul} wherein the amplification
of beta decay and the suppression of inverse beta decay will be reviewed.
In Sec.\ref{glue} the attractive gluon exchange potential will be discussed
with regard to enhancement factors for the production of quark anti-quark
pairs, i.e. quark jets. Final state interactions induced by the Higgs field
are discussed in Sec.\ref{higgs} for \begin{math} Z\bar{Z} \end{math} and
\begin{math} W^+W^- \end{math} production. The Higgs effects become more
important as the mass increases. In principle these effects may be of use in
experimental probes which seek to verify that the Higgs field exists.
This point is briefly discussed in the concluding Sec.\ref{ending}.

\section{The Coulomb Potential\label{coul}}

Consider the inverse beta decay of a nucleus written as the reaction
\begin{equation}
\bar{\nu}_e+(Z+1,A)\to (Z,A)+e^+ .
\label{coul1}
\end{equation}
The finally produced positron interacts with the final nucleus via the
repulsive Coulomb potential
\begin{equation}
U_+(r)=\frac{\hbar c\alpha Z}{r}.
\label{coul2}
\end{equation}
Since the nucleus is much more massive than is the positron, it is normally
sufficient to treat the Coulomb interaction potential as if it were
{\em external}. The positron energy equation then reads
\begin{equation}
\big(E-U_+(r)\big)^2=m^2c^4+c^2|{\bf p}|^2.
\label{coul3}
\end{equation}
Relativistic Hamilton-Jacobi dynamics assert that the momentum is the
gradient of the positron action
\begin{equation}
{\bf p}={\bf grad}W({\bf r},E).
\label{coul4}
\end{equation}
The radial solution of Eqs.(\ref{coul3}) and (\ref{coul4}) reads
\begin{eqnarray}
c^2p(r,E)^2&=&c^2\left[\frac{\partial W(r,E)}{\partial r}\right]^2
=\big(E-U_+(r)\big)^2-m^2c^4,
\nonumber \\
c^2p(r,E)^2&=&
\left[E-\frac{\hbar c\alpha Z}{r}+mc^2\right]
\left[E-\frac{\hbar c\alpha Z}{r}-mc^2\right].
\label{coul5}
\end{eqnarray}
The classically allowed \begin{math} (p^2>0) \end{math} and
classically disallowed \begin{math} (p^2<0) \end{math} regions
in the radial coordinate \begin{math} r \end{math} are defined by
\begin{eqnarray}
0<r<a \ \ {\rm or} \ \ r>b \ \ &\Longrightarrow & \ \ {\rm (allowed)},
\nonumber \\
a<r<b \ \ &\Longrightarrow & \ \ {\rm (disallowed)},
\label{coul6}
\end{eqnarray}
wherein
\begin{equation}
a=\frac{\hbar c\alpha Z}{E+mc^2}\ \ {\rm and}
\ \ b=\frac{\hbar c\alpha Z}{E-mc^2}\ .
\label{coul7}
\end{equation}
The reaction suppression is described by the barrier factor
\begin{math} B \end{math} for the regime in which classical
motion is forbidden; In detail
\begin{eqnarray}
B &=& \frac{2}{\hbar }{\Im }m\big|W(b,E)-W(a,E)\big|=
\frac{2}{\hbar }\int_a^b \big|{\Im }m [p(r,E)]\big|dr,
\nonumber \\
B &=& \frac{2}{\hbar c}
\int_a^b\sqrt{\left|E-\frac{\hbar c\alpha Z}{r}+mc^2\right|
\left|E-\frac{\hbar c\alpha Z}{r}-mc^2\right|}\ dr,
\nonumber \\
B(E,Z\alpha ) &=& 2\pi Z\alpha
\left[\frac{E}{\sqrt{E^2-m^2c^4}}-1\right]=
2\pi Z\alpha
\left[\left(\frac{c}{v}\right)-1\right],
\label{coul8}
\end{eqnarray}
where \begin{math} v \end{math} is the positron velocity. In the
non-relativistic limit \begin{math} v<<c \end{math}, the Coulomb
barrier factor \begin{math} B\approx (2\pi Z\alpha c/v) \end{math} is
well known. Eq.(\ref{coul8}) represents the relativistic theory in
which the barrier factor vanishes in the high energy limit
\begin{math} (v\to c)  \end{math}.

The physical picture in the relativistic theory is worthy of note. The
``tunnelling'' through the barrier is in reality electronic ``pair creation''
under the barrier for \begin{math} (a<r<b) \end{math}. When the pair is
created the positron half of the pair rushes off to infinity
\begin{math} (b<r<\infty ) \end{math}. The electron half of the pair falls into
the center \begin{math} (0<r<a) \end{math} converting one of the nuclear protons
into a neutron and emitting an electron neutrino. The total inverse beta decay
reaction may then be represented as
\begin{eqnarray}
({\rm vacuum})&\to & e^- +e^+,
\nonumber \\
\bar{\nu}_e+e^- +(Z+1,A)&\to & (Z,A),
\label{coul9}
\end{eqnarray}
for which Eq.(\ref{coul1}) is the {\em total} reaction.
The full suppression factor cross section ratio
induced by the Coulomb repulsion between the positron and the
final state nucleus is given by
\begin{eqnarray}
S(E,Z) &=&
\frac{\sigma \Big[\bar{\nu}_e+(Z+1,A)\to (Z,A)+e^+ \Big]}
{\sigma^{(0)}\Big[\bar{\nu}_e+(Z+1,A)\to (Z,A)+e^+ \Big]}\ ,
\nonumber \\
S(E,Z) &=&\frac{B(E,Z\alpha )}
{exp\big(B(E,Z\alpha )\big)-1}\ .
\label{coul10}
\end{eqnarray}
Eq.(\ref{coul10}) concludes our discussion for the case
of inverse beta decay.


\begin{figure}[bp]
\epsfxsize=6.0cm \centerline{\epsffile{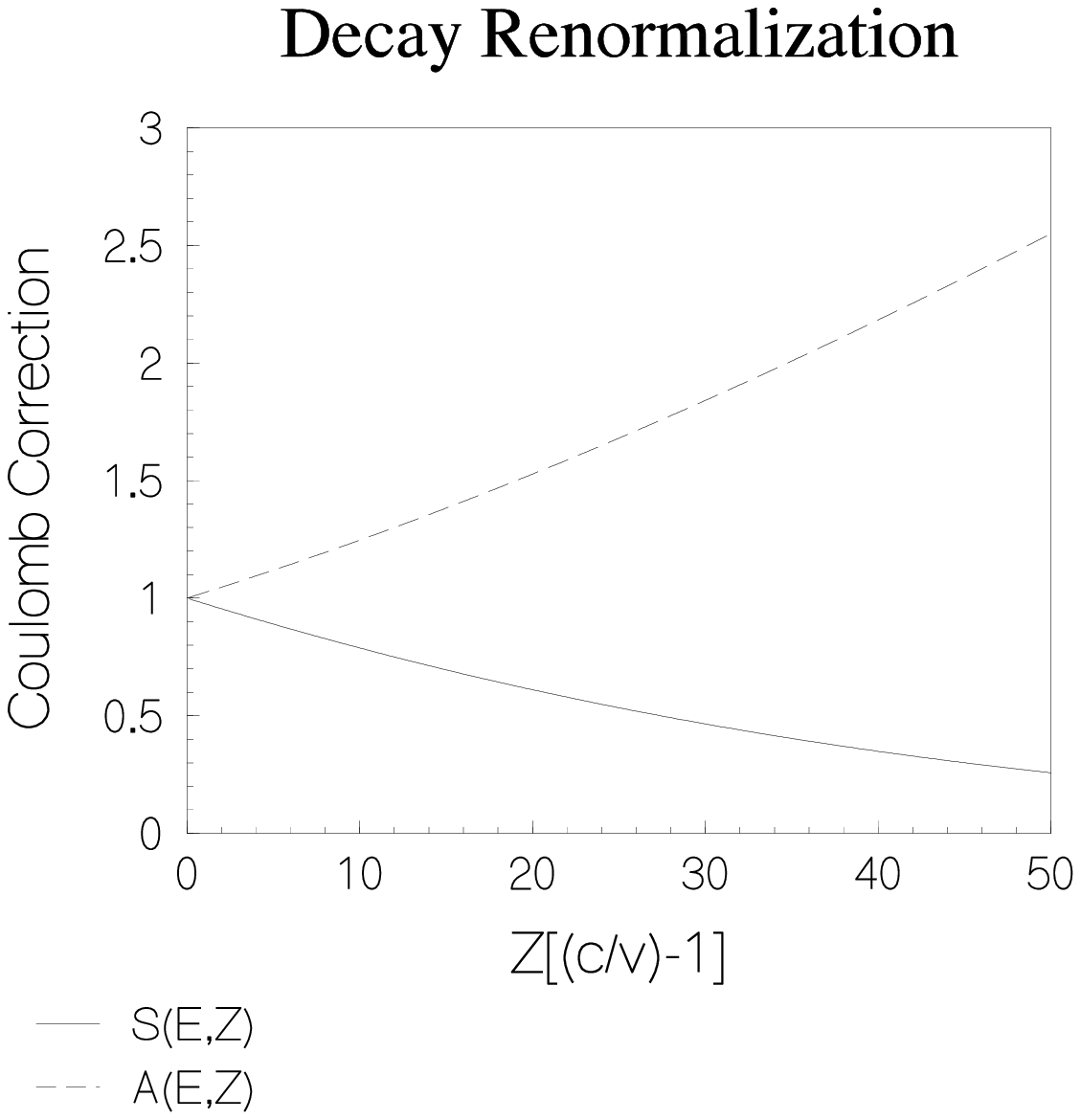}}
\caption{For an outgoing beta decay electron or inverse beta decay
positron with energy $E=\{mc^2/\sqrt{1-(v/c)^2}\}$ there will be,
respectively, an attraction or repulsion from the central nuclear final state
charge $Ze$. Shown are the curves for the electron rate amplification
$A(Z,E)$ and the positron rate suppression $S(Z,E)$ implicit
in the conventional Coulomb final state corrections.
\label{hugfig1}}
\end{figure}


For the case of beta decay
\begin{equation}
(Z-1,A)\to (Z,A)+e^- +\bar{\nu}_e ,
\label{coul11}
\end{equation}
the Coulomb potential between the outgoing electron and the nucleus
is attractive
\begin{equation}
U_-(r)=-\frac{\hbar c\alpha Z}{r}\ .
\label{coul12}
\end{equation}
The Hamilton-Jacobi equation for the attractive Coulomb energy
reads
\begin{equation}
\big(E-U_-(r)\big)^2=m^2c^4+c^2|{\bf p}|^2
\ \ {\rm wherein}\ \ {\bf p}={\bf grad}W(E,{\bf r}).
\label{coul13}
\end{equation}
Since there is a particle anti-particle ``duality'' corresponding
to positive and negative energy solutions in any relativistic theory, if an
electron sees an attractive potential then the positron will see a repulsive
potential. Relativistic dynamics with Poincar\'e symmetry automatically includes
both particle and antiparticle dynamics. Employing this duality of solutions
one finds that the beta decay for the electron is again described by
Eq.(\ref{coul8}) but this time with an amplification factor. The full ratio
of decay rates corresponds to
\begin{eqnarray}
A(E,Z) &=&
\frac{\Gamma \Big[(Z-1,A)\to (Z,A)+e^- +\bar{\nu}_e \Big]}
{\Gamma^{(0)} \Big[(Z-1,A)\to (Z,A)+e^- +\bar{\nu}_e \Big]}\ ,
\nonumber \\
A(E,Z) &=&\frac{B(E,Z\alpha )}
{1-exp\big(-B(E,Z\alpha )\big)}\ .
\label{coul14}
\end{eqnarray}

The suppression factor for an outgoing positron and the amplification factor
for an outgoing electron are plotted in Figure \ref{hugfig1}. For the
inverse beta decay of Eq.(\ref{coul10}), the positron emerges with velocity
\begin{equation}
v=\frac{c\sqrt{E^2-m^2c^4}}{E}
\label{coul15}
\end{equation}
and the cross section is suppressed by the coulomb interaction factor $S$.
For the beta decay case in Eq.(\ref{coul14}), the electron can still emerge with
the velocity in Eq.(\ref{coul15}) but the decay rate is enhanced with an
amplification factor $A$.

\section{The Gluon Exchange Potential\label{glue}}

Consider the production of a quark and an anti-quark with momenta
\begin{math} p \end{math} and \begin{math} \bar{p} \end{math}.
The pair interacts with an attractive gluon exchange potential
\begin{math} U_{\bar{q}q}(r) \end{math}. On a short distance scale
one expects a Coulomb-like potential with a strong interaction charge
which dominates the actual Coulomb potential; i.e.
\begin{equation}
U_{\bar{q}q}(r)=-\frac{4}{3}\left(\frac{g^2}{4\pi \epsilon_0 r}\right)
=-\frac{4}{3}\left(\frac{\hbar c \alpha_s}{r}\right)
\ \ {\rm as}\ \ r\to 0.
\label{glue1}
\end{equation}
On a larger distance scale, the potential is discussed
in \ref{qp}.


\begin{figure}[bp]
\centerline{\epsfxsize=6.0cm \epsfbox{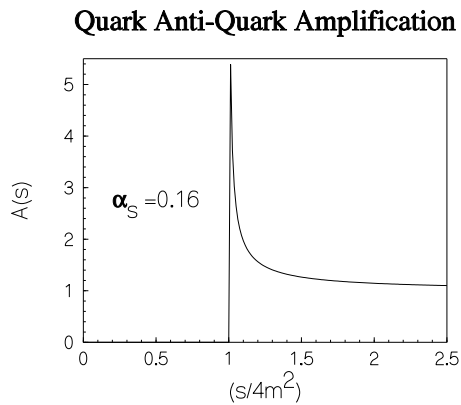}}
\caption{The gluon exchange potential amplification of quark anti-quark jet
production is plotted as a function of the invariant mass squared.
The amplification begins at threshold. A reasonable but approximate
value for the strong coupling strength $\alpha_s $ has been employed.
\label{hugfig2}}
\end{figure}


The total mass
\begin{math} \sqrt{s}  \end{math} of the final state pair is
determined by
\begin{equation}
c^2 s=-P^2=-(p+\bar{p})^2=2(c^2m^2-\bar{p}\cdot p).
\label{glue2}
\end{equation}
In the center of mass reference frame of the pair
\begin{math} ({\bf P}={\bf p}+\bar{\bf p}=0)  \end{math},
kinematics dictates
\begin{equation}
-c^2\bar{p}\cdot p=\bar{E}E-c^2\bar{\bf p}\cdot {\bf p}
=c^4m^2+2c^2|{\bf p}|^2;
\label{glue3}
\end{equation}
In detail, the relative momentum of the quark anti-quark pair is given by
\begin{equation}
|{\bf p}|=c\sqrt{(s/4)-m^2}\ .
\label{glue4}
\end{equation}
The enhancement factor for the quark anti-quark jet production then
follows a form closely analogous to the Coulomb case in Eqs.(\ref{coul12})
and (\ref{coul14}). The production amplification is
\begin{eqnarray}
A_{\bar{q}q}(s) &=&
\frac{\Gamma_{\bar{q}q}(s)}
{\Gamma_{\bar{q}q}^{(0)}(s)}\ ,
\nonumber \\
B_{\bar{q}q}(s)&=& \frac{4\pi \alpha_s}{3}\left[\sqrt{\frac{s}{s-4m^2}}
\ -1\right],
\nonumber \\
A_{\bar{q}q}(s) &=&\frac{B_{\bar{q}q}(s)}
{1-exp\big(-B_{\bar{q}q}(s)\big)}\ ,
\label{glue5}
\end{eqnarray}
which has been plotted in Figure \ref{hugfig2}. The amplification
is particularly strong near the threshold value
\begin{math} s_0=4m^2  \end{math}.

\section{The Higgs Exchange Potential\label{higgs}}

\begin{figure}[bp]
\centerline{\epsfxsize=6.0cm \epsfbox{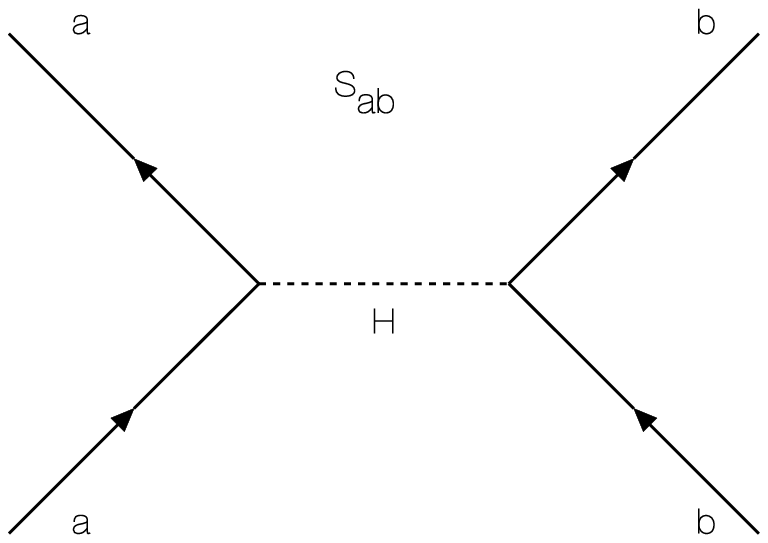}}
\caption{The exchange of a
Higgs boson between two particles gives rise to the attractive
potential $U_{ab}=-(\sqrt{2}/4\pi)(G_FM_aM_b/r)\exp(-M_Hr/\hbar c)$. The
action $S_{ab}$ of the exchange is examined in detail.
\label{hugfig3}}
\end{figure}

The calculation of Higgs exchange amplification factor from the
potential in Eq.(\ref{intro4}) is a bit more delicate due to
the screening effect of the Higgs mass \begin{math} M_H \end{math}.
As shown in what follows, it turns out that the Higgs mass drops out of the
result since the amplification factor is determined by the wave
function of the two produced particle at zero distance for a fixed
time. In effect, this represents a ``zero space time interval''
for the exchange and it is well known that the nature of the light
cone singularity in the mass propagator is mass independent. The
Higgs boson exchange Feynman diagram producing the exchange
potential is shown in Figure \ref{hugfig3}.


The action associated with this exchange is given by
\begin{equation}
S_{Higgs}=\frac{\sqrt{2}G_F}{2 c^5}
\int \int T(x)D(x-y)T(y)d^4x d^4y,
\label{higgs1}
\end{equation}
wherein \begin{math} T(x) \end{math} is the trace of the stress tensor
and \begin{math} D(x-y) \end{math} is the Higgs boson propagator
\begin{equation}
D(x-y)=\hbar^2\int\left[
\frac{e^{ip\cdot(x-y)/\hbar}}{p^2+(M_Hc)^2-i0^+}\right]
\frac{d^4p}{(2\pi \hbar )^4}.
\label{higgs2}
\end{equation}
A more physical space-time representation of the Higgs boson propagation
follows from the Schwinger proper time representation
\begin{equation}
D(x-y)=\frac{M_H}{ 8\pi^2\hbar }\int_0^\infty
e^{[iM_H/2\hbar ]\{-c^2\tau +[(x-y)^2/\tau ]\}}
\left(\frac{d\tau }{\tau^2 }\right).
\label{higgs3}
\end{equation}
For two particles moving at uniform velocities the trace of the stress
tensor quasi-classical sources reads
\begin{math}T_{a,b}(x)=-M_{a,b}c^3\int \delta (x-v_{a,b}\tau)d\tau\end{math}.
Eq.(\ref{higgs1}) now yields the action
\begin{equation}
S_{ab}=\left({\sqrt{2} G_F M_a M_b \over c }\right)
c^2\int_{-\infty}^\infty \int_{-\infty}^\infty
D(v_a \tau_a -v_b \tau_b)d\tau_a d\tau_b\ .
\label{higgs4}
\end{equation}
If Eq.(\ref{higgs3}) is substituted into Eq.(\ref{higgs4}), then the
resulting Gaussian integrals over
\begin{math} d\tau_a \end{math} and \begin{math} d\tau_b \end{math}
can be performed yielding
\begin{equation}
S_{ab}=\left(\frac{\sqrt{2} G_F m_a m_b }{ c }\right)
\int_0^\infty \tilde{F}(v_a,v_b,\tau )\left(\frac{d\tau }{ \tau}\right),
\label{higgs5}
\end{equation}
wherein
\begin{equation}
\tilde{F}(v_a,v_b,\tau )=
\left({ c^2\over 4\pi \sqrt{(v_a\cdot v_b)^2-c^4}}\right)
e^{-iM_Hc^2\tau /2\hbar }.
\label{higgs6}
\end{equation}
The Higgs mass \begin{math} M_H  \end{math} drops out of the final
expression for the imaginary part of the action,
\begin{equation}
{\Im }m\ S_{ab}=-\left({\sqrt{2} G_F M_a M_b \over 8 c }\right)
\left( {M_aM_b c^2\over \sqrt{(p_a\cdot p_b)^2-(M_aM_b c^2)^2} }\right),
\label{higgs7}
\end{equation}
wherein the momenta \begin{math} p_a=M_av_a \end{math} and
\begin{math} p_b=M_bv_b \end{math} have been introduced.

Suppose the production of a particle anti-particle pair each of mass
\begin{math} M \end{math}. Associated with such a mass is a
weak coupling strength
\begin{equation}
\alpha_F(M)=\left(\frac{\sqrt{2}G_FM^2}{4\pi \hbar c}\right)
\label{higgs8}
\end{equation}
such that
\begin{equation}
B_{pair}(s)=-\frac{2}{\hbar}{\Im}mS_{pair}
=2\pi \alpha_F(M)\left(\frac{M^2}{\sqrt{s(s-4M^2)}}\right).
\label{higgs9}
\end{equation}
The resulting Higgs induced amplification factor is determined by
\begin{equation}
A_{pair}(s)=\frac{B_{pair}(s)}{1-\exp(-B_{pair}(s))}\ .
\label{higgs10}
\end{equation}
In this regard one may consider the reactions
\begin{eqnarray}
e^+ + e^-&\to & W^+ + W^-,
\nonumber \\
e^+ + e^-&\to & Z+\bar{Z}.
\label{higgs11}
\end{eqnarray}
The amplification coupling strengths for the above reactions
are, respectively,
\begin{eqnarray}
2\pi \alpha_F(M_W)& \approx & 0.0532,
\nonumber \\
2\pi \alpha_F(M_Z) & \approx & 0.0687.
\label{higgs12}
\end{eqnarray}
For these massive particles the Higgs boson exchange
induced amplification is somewhat larger than the photon
exchange amplification which contributes in the
\begin{math} W^+W^- \end{math} production case.


\begin{figure}[tp]
\centerline{\epsfxsize=6.0cm \epsfbox{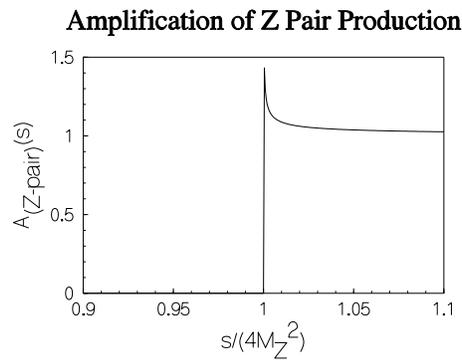}}
\caption{Shown is the amplification factor $A_{\rm (Z-pair)}(s)$ of the
$Z$ pair production reaction
$e^++e^-\to Z+\bar{Z}$ due to a Higgs boson exchange.
\label{hugfig4}}
\end{figure}


In Figure \ref{hugfig4}, we exhibit the amplification factor for
\begin{math} Z\bar{Z} \end{math} production due to the exchange
potential of the Higgs boson; It is
\begin{eqnarray}
A_{\rm (Z-pair)}(s)&=&\frac{\Gamma(e^++e^-\to Z+\bar{Z})}
{\Gamma^{(0)}(e^++e^-\to Z+\bar{Z})}\ ,
\nonumber \\
A_{\rm (Z-pair)}(s)&=&\frac{B_{\rm (Z-pair)}(s)}
{1-\exp[-B_{\rm (Z-pair)}(s)]}\ .
\nonumber \\
B_{\rm (Z-pair)}(s)&=&
2\pi \alpha_F(M_Z)\left(\frac{M_Z^2}{\sqrt{s(s-4M_Z^2)}}\right),
\label{higgs13}
\end{eqnarray}


\begin{figure}[tp]
\centerline{\epsfxsize=6.0cm \epsfbox{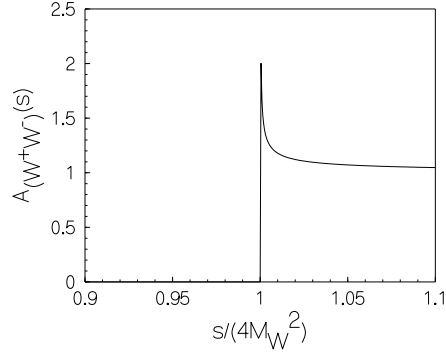}}
\caption{Shown is the amplification factor $A_{W^+W^-}(s)$ of the
$W^+W^-$ pair production reaction $e^++e^-\to W^+ + W^-$ due to
{\it both} Higgs boson exchange {\it and} photon exchange. Both
Higgs exchange and photon exchange contribute to the amplification
factor yielding a somewhat larger effect than for the case of
$e^++e^-\to Z+\bar{Z}$. \label{hugfig5}}
\end{figure}


The case of \begin{math} W^+W^-\end{math} production, the enhancement is
due to both photon exchange (which surely exists) and Higgs boson exchange
(which may exist). The complete answer for \begin{math} W^+W^-\end{math}
amplified production reads
\begin{eqnarray}
A_{\rm (W-pair)}(s)&=&\frac{\Gamma(e^++e^-\to W^+ +W^-)}
{\Gamma^{(0)}{(e^++e^-\to W^+ +W^-)}}\ ,
\nonumber \\
A_{\rm (W-pair)}(s)&=&\frac{B_{\rm (W-pair)}(s)}
{1-\exp[-B_{\rm (W-pair)}(s)]}\ ,
\nonumber \\
B_{\rm (W-pair)}(s)&=&
2\pi \alpha_F(M_W)\left(\frac{M_W^2}{\sqrt{s(s-4M_W^2)}}\right)
\nonumber \\
&\ &\ \ \ \ \ \ \ \ \ \
+\pi \alpha \left(\frac{\sqrt{s}-\sqrt{s-4M_W^2}}{\sqrt{(s-4M_W^2)}}\right),
\label{higgs14}
\end{eqnarray}
which is plotted in Figure \ref{hugfig5}. The amplification factor for
\begin{math} W^+W^- \end{math} production is more pronounced than the
amplification factor for \begin{math} Z\bar{Z} \end{math} production
since photon exchange contributes to the former process but not
contribute to the later.

\section{Conclusions\label{ending}}

The threshold amplification and/or suppression factors familiar
from the theory of final state interactions has been applied in this
work in a higher energy regime. In particular we have considered
final state interactions involving the Higgs boson under the supposition
that it exists. Even below the threshold for the physically real
Higgs particle  production, the Higgs field can act as a messenger field
entering into enhanced production rates for pairs of heavy particles
such as \begin{math} Z\bar{Z}  \end{math}, \begin{math} W^+W^- \end{math}
or \begin{math} t\bar{t}  \end{math} pairs\cite{9}.
The sharp peaks shown in the plots of enhancement factors will be considerably
``rounded'' due to (i) particle lifetime effects, (ii) radiative corrections
and (iii) energy resolution factors from the energy distributions in incoming
beams. Nevertheless, even if a sharp peak no longer appears, the physically
``smoothed'' threshold regime will be shifted. Since the production amplification
is above the threshold mass squared,
i.e. \begin{math} s>s_0\equiv 4M^2  \end{math}, it follows that the threshold
transition region will occur at a mass slightly higher that the threshold to
be expected if the amplification were ignored. For example, experimental
reaction threshold mass shifts of order
\begin{eqnarray}
e^++e^- \to  Z+\bar{Z} \ \ \ &\Rightarrow &
\ \ \ \Delta M_Z \approx M_Z\alpha_F(M_Z),
\nonumber \\
e^++e^- \to  W^+ + W^- \ \ \ &\Rightarrow &
\ \ \ \Delta M_W \approx M_W[\alpha_F(M_W)+0.5\alpha ],
\label{ending1}
\end{eqnarray}
would not be unreasonable and might constitute an unexpected probe of
the Higgs field existence.

\appendix

\section{Quark Potentials \label{qp}}

The one gluon exchange potential between a quark and anti-quark
has been approximated as
\begin{equation}
V_{Glue}(r)=
\left(\frac{\hbar c\alpha_s }{r}\right)({\bf T}_1\cdot {\bf T}_2)
=\int \left(\frac{4\pi \hbar c\alpha_s }{|{\bf k}|^2}\right)
e^{i{\bf k\cdot r}}\frac{d^3 {\bf k}}{(2\pi )^3}
({\bf T}_1\cdot {\bf T}_2).
\label{qp1}
\end{equation}
In reality, the strong interaction coupling strength itself depends
on \begin{math} |{\bf k}|^2  \end{math} so that the Coulomb-like
potential is modified to read
\begin{equation}
\tilde{V}_{Glue}(r)=4\pi \hbar c
\int \left(\frac{\alpha_s(|{\bf k}|^2) }{|{\bf k}|^2}\right)
e^{i{\bf k\cdot r}}\frac{d^3 {\bf k}}{(2\pi )^3}
({\bf T}_1\cdot {\bf T}_2).
\label{qp2}
\end{equation}
More simply,
\begin{eqnarray}
\tilde{V}_{Glue}(r) &=&
\left[\frac{\hbar c({\bf T}_1\cdot {\bf T}_2)}{r}\right]
\chi (r),
\nonumber \\
\chi(r) &=& \frac{2}{\pi }\int_0^\infty \alpha_s(k^2)sin(k r)
\frac{d k}{k}\ .
\label{qp3}
\end{eqnarray}
If \begin{math} \alpha_s(k^2) \end{math} were a constant, then
Eqs.(\ref{qp3}) would reduce to Eq.(\ref{intro3}). However the Coulomb-like
law from gluon exchange breaks down at large distances.

To see what happens as \begin{math} r\to \infty  \end{math}, one may
presume a finite limit in the form
\begin{equation}
\lim_{k^2\to 0^+}\{\hbar c k^2\alpha_s(k^2)\}=2\sigma ,
\label{qp4}
\end{equation}
and differentiate Eq.(\ref{qp3}) twice with respect to
\begin{math} r \end{math}; i.e.
\begin{eqnarray}
\chi ^{\prime \prime }(r) &=&
-\frac{2}{\pi }\int_0^\infty k \alpha_s(k^2)sin(k r)d k,
\nonumber \\
\lim_{r\to \infty}\chi ^{\prime \prime }(r)
&=& -\frac{2\sigma }{\hbar c}\ .
\label{qp5}
\end{eqnarray}
What is called a ``QCD motivated potential'' results from the
assertion that
\begin{math} \chi ^{\prime \prime }(r) = -(2\sigma /\hbar c)\end{math}
for all of the important distance scales. If this is indeed the case, then
\begin{equation}
\tilde{V}_{Glue}(r) ={\bf T}_1\cdot {\bf T}_2
\left\{ \frac{\hbar c\alpha_s}{r}- \sigma r\right\},
\label{qp6}
\end{equation}
wherein the long range linear part of the potential describes the
intrinsic tension \begin{math} \sigma  \end{math} in a QCD string.
In detail, for the quark anti-quark potential
\begin{equation}
U_{\bar{q}q}(r)=-\frac{4}{3}\left(\frac{\hbar c\alpha_s }{r}\right)
+\tau_{\bar{q}q} r\ \ {\rm where}\ \ \tau_{\bar{q}q} =\frac{4\sigma }{3}\ ,
\label{qp7}
\end{equation}
and for the quark-quark potential
\begin{equation}
U_{qq}(r)=-\frac{2}{3}\left(\frac{\hbar c\alpha_s }{r}\right)
+\tau_{qq} r\ \ {\rm where}\ \ \tau_{qq} =\frac{2\sigma }{3}\ .
\label{qp8}
\end{equation}

\end{document}